\documentclass[]{elsarticle}

\usepackage{hyperref}

\journal{Computer Networks}

\usepackage[yyyymmdd,hhmmss]{datetime}
\usepackage{amssymb,bm,bbm}
\usepackage{graphicx}
\usepackage{eurosym}
\usepackage{euscript}

\newcommand*\diff{\mathop{}\!\mathrm{d}}
\newcommand{\equaref}[1]{(\ref{eq:#1})}

\usepackage[normalem]{ulem}
\usepackage{url}
\usepackage{xcolor}

\usepackage{amsmath}

\usepackage{amsthm}
\theoremstyle{plain}
\newtheorem{thm}{Theorem}[section]

\newtheorem{prop}[thm]{Proposition}

\theoremstyle{definition}

\theoremstyle{remark}
\newtheorem{rem}{Remark}

\usepackage{algpseudocode}

\newcommand{\x}{\mathbf x}

\newcommand{\sX}{\mathcal X}

\newcommand{\sS}{\mathcal S}

\newcommand{\sB}{\mathcal B}
\newcommand{\sA}{\mathcal A}

\newcommand{\expC}{\EuScript C}

\newcommand{\greedy}{\textproc{Greedy}}
\newcommand{\localswap}{\textproc{LocalSwap}}

\newcommand{\lru}{\textproc{LRU}}

\newcommand{\duel}{\textproc{Duel}}
\newcommand{\netduel}{\textproc{NetDuel}}

\DeclareMathOperator*{\argmin}{arg\,min}

\usepackage{mathtools}

\usepackage{amsmath}

\usepackage{mleftright}
\usepackage{amssymb}

\usepackage{epstopdf}
\usepackage{verbatim}
\usepackage{enumitem}

\begin{document}

\begin{frontmatter}

\title{Content Placement in Networks of Similarity Caches} 


\author[unito]{Michele Garetto\corref{mycorrespondingauthor}}
\cortext[mycorrespondingauthor]{Corresponding author}
\ead{michele.garetto@unito.it}

\author[polito]{Emilio Leonardi}
\ead{emilio.leonardi@polito.it}

\author[inria]{Giovanni Neglia}
\ead{giovanni.neglia@inria.fr}

\address[unito]{Universit\`{a} degli Studi di Torino, C.so Svizzera 185, Torino, Italy}
\address[polito]{Politecnico di Torino, C.so Duca degli Abruzzi 24, Torino, Italy}
\address[inria]{Inria - Universit\'e C\^ote d'Azur, 2004 route des Lucioles, Sophia Antipolis, France}

\begin{abstract}
Similarity caching systems have recently attracted the attention of the scientific community, as they 
can be profitably used in many application contexts, like multimedia retrieval, advertising, object recognition, recommender systems
and online content-match applications. In such systems, a user request for an object $o$, which is not in 
the cache, can be (partially) satisfied by a similar stored object $o$', at the cost of a loss of user utility.
In this paper we make a first step into the novel area of similarity caching networks,
where requests can be forwarded along a path of caches to get the best efficiency-accuracy tradeoff.
The offline problem of content placement can be easily shown to be NP-hard, while different
polynomial algorithms can be devised to approach the optimal solution in discrete cases.
As the content space grows large, we propose a continuous problem formulation whose solution 
exhibits a simple structure in a class of tree topologies.
We verify our findings using synthetic and realistic request traces. 
\end{abstract}


\end{frontmatter}


\section{Introduction}
Similarity caching is an extension to traditional (exact) caching, whereby a request for an object
can be satisfied by providing a similar cached item, under a dissimilarity cost.
In some cases, user requests are themselves queries for objects similar to a given one
(similarity searching~\cite{gionis99}). Caching at network edges can drastically reduce the latency experienced by users, 
as well as backbone traffic and server provisioning.

{\color{black}
Similarity searching and caching have several applications 
in multimedia retrieval~\cite{falchi08}, contextual advertising~\cite{pandey09}, object recognition~\cite{drolia2017cachier,drolia2017precog,guo2018foggycache,venugopal2018shadow}, recommender systems~\cite{pandey09,sermpezis18}, 
online prediction serving systems \cite{crankshaw17}.
However, theoretical understanding of similarity caching and development of 
related algorithms and policies are at their early stages. 

Simple modifications to the Least Recently Used 
policy (\lru{}) which deal with approximate (soft) hits were proposed in \cite{falchi08,pandey09}.
In~\cite{sermpezis18} authors have studied how to statically place contents 
in edge caches of a cellular network, given their popularities and the 
utility for a user interested in content $o$ to receive a similar content $o'$. 
An adversarial setting was studied in \cite{chierichetti09} by competitive analysis.
The authors of~\cite{sabnis21} have proposed a similarity caching policy tailored for the case 
when cached objects may be embedded in $\mathbb R^d$ with a distance that captures dissimilarity costs. 
The work most closely related to this paper is \cite{infocom2020}, where we have 
analyzed a single similarity cache in the offline, adversarial, and stochastic settings,
proposing also some dynamic online policies to manage the cache. 

We mention that many researchers have studied networks of exact caches~(e.g., \cite{rosensweig13,shanmugan13,choungmo14,ioannidis18,leonardi18,li20}),
however their results cannot be applied to the similarity caching setting, which is a fundamentally different problem 
(in exact caching there is no notion of distance between objects). 
To the best of our knowledge, only the recent letter~\cite{zhou20} has considered a network of similarity caches, 
where requests can be forwarded along a path of caches towards a repository storing all objects, at the cost of increasing
delays and resource consumption. The authors of~\cite{zhou20} have proposed a heuristic based on the gradient descent/ascent algorithm to jointly decide request routing and caching, similarly to what done in~\cite{ioannidis18} for exact caches but without the corresponding theoretical guarantees. 
The proposed algorithm requires memory proportional to the size of the catalog, and appears to be computationally feasible only on 
small-scale systems. 
    
In our work, similarly to ~\cite{zhou20},
we focus mainly on the offline setting, i.e., the problem of statically placing objects in the caches so as to minimize the 
expected cost under known content request rates and routing.   
In contrast to ~\cite{zhou20}, we first propose algorithms with guaranteed performance, 
and then we move to the continuous limit of the large requests/catalog space, where we investigate
the {\em structure} of the optimal solution. 
} 

Our contributions are the following:
\begin{enumerate}
	\item while the content placement problem in networks of similarity caches is NP-hard, 
	we show that it can be formulated as the maximization of a sub-modular function  over a matroid; therefore a polynomial 
	\greedy{} algorithm can be defined with $1/2$ approximation ratio;
	\item we propose the randomized \localswap{} algorithm that does not enjoy worst-case guarantees as \greedy{}, but asymptotically converges to a locally optimal solution;   
	\item we characterize the structure of the optimal similarity-caching placement problem in special cases; in particular, we show that, under mild assumptions,
	when the cache network has a regular tree structure and requests arrive only at the leaves the optimal solution  
	in the large catalog regime has a relatively simple structure;
	\item we show that the above structure is lost in general networks, analyzing a simple tandem network where requests 
	arrive at both caches;
	\item we propose an online, $\lambda$-unaware policy called \netduel, that extends \duel{} \cite{infocom2020} to the networked setting;
	\item we illustrate our findings considering both synthetic and real request processes for Amazon items.
\end{enumerate}

\section{Main assumptions and  problem formulation}
\label{s:model}
Let $\sX$ be the (finite or infinite) set of objects that can be requested by the users.  
We assume that all objects have equal size and cache $i$ can store up to $k_i$ objects. 

We consider a network of caches with requests potentially arriving at every node.
Some nodes can act as content repositories, where (a subset of) requests can be satisfied exactly or with a small approximation cost.
Specifically, we assume that each request has at least one repository acting as \lq authoritative server' for it,
meaning that the approximation cost at the content repository is either zero or it is negligible as compared to the fixed cost to reach 
the repository (see next). Let $\mathcal{K}$ be the set of all nodes in the network (including caches and repositories).

A request $r$ is a pair $(o,i)$ where $o$ is the requested object and $i$ is the node where the request first enters the network. 
Every request is issued according to a  Poisson process with rate $\lambda_r$.

At each cache, for any two objects $x$ and $y$ in $\sX$ there is a non-negative (potentially infinite) cost $C_a(x,y)$ to locally 
approximate $x$ with $y$. We consider $C_a(x,x)=0$. We assume that caches can efficiently compute, upon arrival of a request for $x$, the closest
stored object $y$. This is typically done resorting to locality sensitive hashing (LSH)~\cite{pandey09}. 

Moreover, there is an additional retrieval cost $h(i,j)$ to reach node $j$ from cache $i$, which is assumed to
increase as more and more hops need to be traversed by the request. Costs $h(i,j)$ represent the additional penalty 
(in terms of network delay) incurred by requests, in addition to the approximation cost $C_a$. If a request from $i$ cannot be forwarded to cache $j$, then $h(i,j)= +\infty$.

We call an approximizer $\alpha$ a pair $(o',j)$, where object $o'$ has been placed at cache $j$. 
If a request $r=(o,i)$ is served by object $o'$ at node $j$, it will incur a total cost $C(r,\alpha) = C_a(o,o') + h(i,j)$, 
that depends on how dissimilar $o$ is from $o'$ and how far node $i$ is from node $j$. For approximizers located at a content repository $j$, we take
$C(r,\alpha) = h(i,j)$, neglecting the local approximation cost.

We assume that each cache knows how to route each request to a corresponding repository. Nevertheless, deciding if a request should be served locally or should be forwarded along the path to the repository is still a challenging problem to solve in a distributed way: while a relatively good approximizer can be found at a cache $i$, a better one may be located at an upstream cache $j$, justifying the additional cost $h(i,j)$.
This is
in sharp contrast to what happens
in exact caching network, where the forwarding operation is straightforward (a request is forwarded upon a miss).

In our initial investigation, we will suppose that optimal forwarding strategy is   available
at all caches, i.e., that each cache knows whether to solve a request locally or forward it towards the repository.
This assumption is reasonable in two possible scenarios: i) when caches exchange meta-data information
about their stored objects (this is acceptable when content is static or quasi-static); 
ii) when the dominant component of the delay is content download, so that,
prior to download, small request messages can go all the way up to the repository and back, dynamically finding the best
approximizer along the path. We leave to future work the challenging case in which optimal forwarding is not available 
at the nodes.  

A consequence of our assumptions is that each request $r$ will be served minimizing the total cost, i.e.,
given $\sS$ the initial set of approximizers  at content repositories,  and $\mathcal A$ the set of approximizers at the caches, we have
\begin{equation}
C(r,\mathcal A) = \min_{\alpha \in \mathcal A \cup \mathcal S} C(r,\alpha). 
\end{equation}

In what follows we will consider two main instances for $\sX$ and $C_a()$.
In the first instance,  $\sX$ is a finite set of objects and thus the approximation cost can be characterized by an 
$|\sX| \times |\sX|$ matrix of non-negative values.
This case could well describe the (dis)similarity of contents (e.g.~videos) in a finite catalog.
In the second instance, $\sX$ is a subset of $\mathbb R^p$ and $C_a(x,y)=f(d(x,y))$, where $f: \mathbb R^+ \to \mathbb R^+$ 
is a non-decreasing non-negative function and $d(x,y)$ is a metric in $\mathbb R^p$ (e.g.~the Euclidean one). 
This case is more suitable to describe objects characterized by continuous features, as in 
machine learning applications. For example, consider a query to retrieve similar images, as one can issue to
\url{images.google.com}. The set of images the user may query Google for is essentially unbounded, 
and in any case it is larger than the catalog of 
images Google has indexed.

In the continuous case, we assume 
a spatial density of requests arriving at each cache defined by a Borel-measurable 
function $\lambda_{x,i}:\,\mathcal{X} \times \mathcal{K} \to \mathbb{R}_+$, i.e., for every Borel set $\mathcal B \subseteq \sX$, 
and every cache $i \in \mathcal{K}$, the rate with which requests for objects in $\mathcal B$ arrive at node $i$ 
is given by $\int_{\mathcal B} \lambda_{x,i} \diff x$.
We will refer to the above two instances as \emph{discrete} and \emph{continuous}, respectively.
 


Under the above assumptions, our goal is to find the optimal static allocation $\sA$ that minimizes the expected cost $\expC(\sA)$ per time unit 
(or per request, if we normalize the aggregate request arrival rate to 1):
\begin{equation}\label{eq:conditionalcost}
\expC(\sA) \triangleq
	\begin{cases}
		\sum_r \lambda_r C(r,\sA), & \textrm{discrete case}\\
		\sum_{i \in \mathcal{K}} \int_{\sX} \lambda_{x,i} C((x,i),\sA) \diff x, & \textrm{continuous case} 
	\end{cases}
\end{equation} 
i.e.,
\begin{equation}
\label{e:static_cost_min}
    \begin{aligned}
    & \underset{\sA}{\text{minimize}} && \expC(\sA)\\
    & \text{subject to} &&\sum_{o: (o,i) \in \sA} 1 \le k_i, \quad \forall i \in \mathcal{K}
    \end{aligned}
\end{equation}

\section{Algorithms for the Discrete case}
In this section, we restrict ourselves to the discrete scenario, as this allows us to make rigorous statements about 
NP-hardness and algorithms' complexity.

\subsection{NP-Hardness and Submodularity}

\begin{prop}
The static off-line similarity caching problem in a network~\eqref{e:static_cost_min} is NP-hard.
\end{prop}
This is an immediate consequence of the fact that, as shown in \cite[Thm.~III.1]{infocom2020}, the static off-line similarity caching 
problem is already NP-hard for a single cache.
Nevertheless,   
we will show in Sec. \ref{sec:cont} that, when the cache network has a regular tree structure,  
a simple characterization of the optimal solution can be determined in the large catalog regime, 
by exploiting a continuous approximation.

Given the initial set  $\sS$ of objects allocated at content repositories, 
we want to pick an additional set $\sA$ of objects and place them at the caches. 
Let $\mathcal I$ denote the set of possible allocations that satisfy cardinality 
constraints at each cache (corresponding to the constraints in \eqref{e:static_cost_min}). 
Let $G(\sA)$ quantify the \emph{caching gain}~\cite{golrezaei12,ioannidis18} from allocation $\sA$ in comparison to the case when each request needs to be served by its 
content repository, i.e. \[G(\sA)= \expC(\emptyset)- \expC(\sA).\]

Problem~\eqref{e:static_cost_min} is equivalent to the following maximization problem
\begin{equation}
\label{e:static_gain_max}
    \begin{aligned}
    & \underset{\sA \in \mathcal I}{\text{maximize}} &&G(\sA).
    \end{aligned}
\end{equation}

\begin{prop}
\label{p:submodularity}
The static off-line similarity caching problem in a network is a submodular maximization problem with matroid constraints.
\end{prop}
The result does not rely on any specific assumption on $C(r,\alpha)$ but for the cost being non-negative. 
In particular, we can define $C(r,\alpha)$ to embed requests' routing constraints. For example, given a request $r=(o,i)$, 
we can enforce the request to be satisfied by the repository of content $o$ or by one of the caches on the 
routing path between node $i$ and the repository (we denote it as $P_{i,o}$). This constraint can be imposed by 
selecting $C((o,i),(o',j)) = \infty $ for each $j \notin P_{i,o}$.
The proof is quite standard and we report it in~\ref{a:proof} for completeness.

\subsection{\greedy{} algorithm and its complexity}
As Problem~\eqref{e:static_gain_max} is the maximization of a monotone non-negative submodular function with matroid constraints, 
the \greedy{} algorithm has $1/2$ guaranteed approximation ratio, i.e., $G(\sA_{\textrm{\greedy}}) \ge \frac{1}{2} \max_{\sA \in \mathcal I} G(\sA)$~\cite{fisher78}. 
We mention that there exists also a randomized algorithm that combines a continuous greedy process and pipage rounding to achieve a $1-1/e$ approximation ratio \emph{in expectation}~\cite{calinescu11}.


The \greedy{} algorithm proceeds from an empty allocation $\sA=\emptyset$ and progressively adds to the current allocation an 
approximizer in $ \arg\!\max_\alpha G(\sA \cup \{\alpha\}) - G(\sA) = \arg\!\max_\alpha \sum_r \lambda_r (C(r,\sA) - C(r, \sA \cup \{\alpha\})) $ 
up to select $\sum_i k_i = K$ objects, where $K$ is the total cache capacity in the network (by respecting local constraints at individual caches). Let $O$, $O_R$, and $N$ denote the number of objects in the catalog, the number of objects that can be requested, and the number of caches in the network.
When choosing the $i$-th approximizer the greedy algorithms needs in general to evaluate $ON-i+1$ possible approximizers, and how they reduce the cost for the set of requests with cardinality at most $O_R N$. The time-complexity of the algorithm is then bounded by $\sum_{i=1}^K O_R N (ON-i+1) =  O_R N ( O N K - K(K-1)/2)$. A smart implementation can avoid to evaluate the gain of all possible approximizers at each step, 
but
despite the optimizations, the \greedy{} algorithm would be too complex for catalogue sizes $O$  beyond a few thousands of objects. 
Moreover, the set of possible requested objects $O_R$ may be much larger than $O$.

\subsection{\localswap{} algorithm and its complexity}
\label{s:lambda_aware}

We now present a different algorithm, called \localswap{}, which is based on the simple idea to systematically 
move to states with a smaller expected cost~\eqref{eq:conditionalcost}. {\localswap{} can be used both in an off-line and on-line scenario.}
It works as follows.  
At the beginning the state of caches is populated by random contents. {\color{black} Then, in the on-line scenario 
	the algorithm adapts the cache state upon every request. In the off-line scenario, instead, a sequence of emulated requests  
	is generated (satisfying the  same statistical  properties of the  original arrival process), and applied to drive cache state changes.}
	 Let $\mathcal{A}_t$ be the allocation
obtained by the algorithm at iteration $t$.
 Upon an (emulated) request $r$ for $o$, \localswap{} computes the  maximum decrement in the expected cost that 
can be obtained by replacing one of the objects currently stored at some cache along the forwarding path with $o$, i.e.,
$\Delta \expC \triangleq \min_{y \in \mathcal{A}_t} \expC(\mathcal{A}_t \cup \{o\} \setminus \{y\})-  \expC(\mathcal{A}_t)$. 
\begin{itemize}
\item if $\Delta \expC < 0$ ($x$ contributes to decrease the cost), then the cache  
replaces $y_e \in \argmin_{y \in \mathcal{A}_t} \expC(\mathcal{A}_t \cup \{o\} \setminus \{y\})$ with $o$;
\item if $\Delta \expC \ge  0$, the cache state is not updated. 
\end{itemize}

{\color{black}
\localswap{} does not  provide worst case guarantees as \greedy, but it asymptotically reaches a locally optimal cache configuration,
defined as a configuration whose cost ~\eqref{eq:conditionalcost} is lower  
than the cost of all configurations that can be obtained by replacing just one content in one cache. On the contrary, \greedy{} does not necessarily reach a local optimal state (as we show below in Sect.~\ref{s:toy_example}).
}
\begin{prop}
\label{p:localswap}
For long enough request sequence \localswap{} converges with probability 1 to a locally optimal cache configuration.
\end{prop}
\localswap{} generalizes a similar algorithm proposed in~\cite{infocom2020} for a single cache (called there ``greedy'') with similar theoretical guarantees. Under the assumption that requests are optimally forwarded, the proof of Proposition~\ref{p:localswap} is essentially the same of \cite[Thm.~V.3]{infocom2020}, so we omit it.
By clever data structure design, the computational cost of each iteration can be kept $\mathcal O(N O_R)$. 
\begin{rem}
Note that  by cascading  \greedy{}  and \localswap{}  it is possible to achieve a locally optimal  cache configuration
  whose approximation ratio is guaranteed to be at least 1/2
 (i.e.,  $G(\sA_{\textrm{\greedy+\localswap}}) \ge \frac{1}{2} \max_{\sA \in \mathcal I} G(\sA)$).
\end{rem}
\subsection{\greedy{}  and \localswap{} in a toy example}
\label{s:toy_example} 
{\color{black} This example shows that 1) \greedy{} does not converge necessarily to a locally optimal cache configuration, and 2) there are both settings  where \greedy{} finds the optimal cache configuration while \localswap{} may not, and settings where \localswap{} finds the optimal cache configuration while \greedy{} does not.}

Consider a scenario with 5 contents $x_i$ for $1\le i \le 5$. Let us assume that $C_a(x_2,x_3)= C_a(x_3,x_4) =0$, $C_a(x_1,x_2)= C_a(x_4,x_5) = \epsilon>0$\footnote{All costs are assumed to be symmetric.}, 
 while $C_a(x_i,x_j)=\infty$ otherwise.  We want to solve the content placement problem for a single cache with $k=2$ and $\lambda_{x_3} > \lambda_{x_2} = \lambda_{x_4} > \lambda_{x_1} = \lambda_{x_5}$. The cost to retrieve the objects from the remote server is $h_s> 2 \epsilon$.
The optimal placement configuration is:  $\{x_2,x_4\}$.
\greedy{} will reach one of the following equivalent sub-optimal configurations $\{x_3,x\}$, with $x\in\{x_1,x_5\}$. 
\localswap{}, on the contrary, will reach the optimal configuration {\color{black} $\{x_2,x_4\}$ (because it is the unique locally optimal configuration). We observe that the configurations reached by \greedy{} are not locally optimal: for example if \greedy{} selects $\{x_3,x_1\}$, it is convenient to replace $x_3$ with $x_4$.}

If we consider two caches $1$ and $2$ in tandem, each of size $k=1$ with requests arriving only to the first cache and retrieval cost equal to $h(1,2)$ if the object is retrieved from cache $2$, and $h(1,2) + h_s$ if it is retrieved by the server. The optimal configurations will maintain a similar structure for $h(1,2)$ small enough. 
In particular the optimal configurations will be: $\{(x_4,1),(x_2,2)\}$ and $\{(x_2,1),(x_4,2)\}$.
\greedy{}  will still reach a state $\{(x_3,1),(x,2)\}$ with $x\in\{x_1,x_5\}$, while \localswap{} will  reach an optimal state.
For $h(1,2)$ large enough the optimal states become $\{(x_3,1),(x,2)\}$ with $x\in\{x_1,x_5\}$  and both previous algorithms will succeed in reaching an optimal solution.
At the same time there are settings for which the configurations $\{(x_3,1),(x_1,2)\}$ and $\{(x_3,1),(x_5,2)\}$ correspond to global minima, the configurations $\{(x_4,1),(x_2,2)\}$ and $\{(x_2,1),(x_4,2)\}$ correspond to local minima, and \greedy{} finds one of the first configurations, while 
 \localswap{} may reach one of the second configurations.
 For example this is the case for $h_s=1$, $h(1,2)= \epsilon = 4/9$, $\lambda_1 = \lambda_5 = 1$, and $\lambda_2 = \lambda_4= 4/3$ and any $\lambda_3>\lambda_2$.

\section{The Continuous case}\label{sec:cont}
When $O_R$ is much larger than $O$, or $O$ is itself very large,
it makes sense to study the request space as continuous. Such continuous representation
permits us to formulate a simplified optimization problem whose solution well approximates
the optimal cost achieved in \emph{discrete} scenarios with large catalog size.   

If the number of objects in the catalog is finite, one could in principle devise a \greedy{} algorithm also for this case, working exactly as in the \emph{discrete} case. Indeed the  problem~\eqref{e:static_cost_min} can be easily shown to be still
 submodular even when requests lies over a  continuous space.
However, one now has to evaluate, for each possible candidate approximizer $\alpha$ to add to the current allocation,
complex integrals over the infinite query space. It is not simple to define in general the complexity of such operations but it 
is evident that previous algorithmic approaches becomes rapidly unfeasible for large set of requests and/or large catalog.

{Hereinafter, we will assume that both the request space and the catalog space are continuous.} 

\subsection{Preliminary: continuous formulation for a single cache} \label{s:back}
As a necessary background, we summarize here some results obtained in \cite{infocom2020} 
for the case of a single cache with capacity $k_1$.
Let $\sB_r(y_0)$ be the closed  ball of radius $r$ around $y_0$, i.e., the set of points $y$ such that 
$d(y ,y_0)\le r$.
The authors of \cite{infocom2020} proved:
\begin{prop}\label{cor2}
Under a homogeneous request process with intensity $\lambda$ over a bounded set $\sX$,
any cache state $\mathcal{A}=\{y_1, \dots, y_{k_1}\}$, such that, for some $r$, the balls $\sB_{r}(y_h)$ for $h=1,\dots, k_1$ are a tessellation of 
$\sX$ (i.e.,~$\cup_h \sB_r(y_h)=\sX$ and $|\sB_r(y_i)\cap \sB_d(y_j)|=0$ for each $i$ and $j$), is optimal.
\end{prop}
Such regular tessellation exists, in all dimensions, under the norm-1 distance, and corresponds to the case
in which balls are squares (assuming that $k_1$ such squares cover exactly the domain $\sX$). 

\begin{figure}
\centering
\includegraphics[scale=0.6]{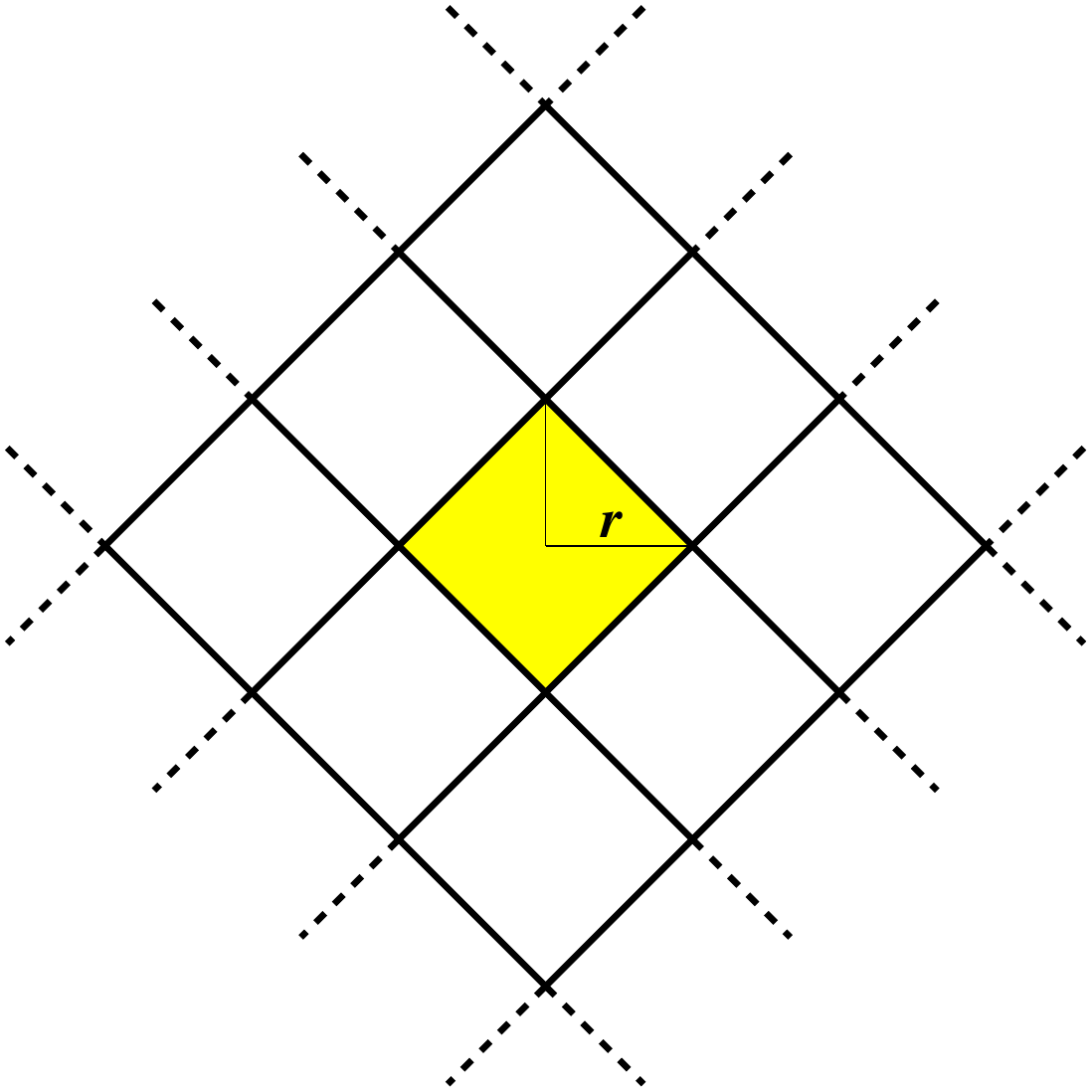}
\caption{Perfect tessellation with square cells in a two-dimensional domain, under the norm-1 distance.}
\label{f:diamonds}
\end{figure}

It is then immediate to analytically compute the optimal cost for this case. For example, 
in a two-dimensional domain (see Fig. \ref{f:diamonds}), requests arriving in a particular ball produce an
approximation cost:
\begin{equation}\label{eq:cr}
c(r) = 4 \int_{0}^{r}\int_{0}^{r-x} (x+y)^\gamma \lambda \diff y \diff x = 4 \lambda \frac{r^{\gamma+2}}{\gamma+2} 
\end{equation}
and the total cost is just $\expC(\mathcal A) = k_1 c(r)$.

 
If the request rate is not space-homogeneous, one can apply the results 
above over small regions $\sX_i$ of $\sX$ where $\lambda_x$ can be approximated by a constant value $\lambda_{\sX_i}$. 
Intuitively, the approximation becomes better and better the more $\lambda_x$
varies smoothly over each Voronoi cell of region $i$. {\color{black} This in particular occurs when $\lambda_x$ is smooth
over the entire domain, and the cache size increases.}
  
Under this approximation, let $k_{i,1}$ be the number of cache 
slots devoted to region $i$ (with the constraint that $\sum_i k_{i,1} =k_1$).
Then, using standard constrained optimization methods, it is possible to determine the optimal value of $k_{i,1}$ as 
function of the local request rate $\lambda_{\sX_i}$.
Without loss of generality, we can assume that domain $\sX$ is partitioned into $M$ regions of unitary area,
on which the request rate is approximately assumed to be constant and equal to $\lambda_i$, $1 \leq i \leq M$.
 
Then, focusing for simplicity on the two dimensional case when $d(x,y)$ is the norm-1,  
and $C_a(x,y)=d(x,y)^\gamma$, each cache slot is used to approximate requests falling in a square of area $1/k_{i,1}$ 
and radius  $r_i = \sqrt{1/(2 k_{i,1})}$. 
Following \equaref{cr}, the approximation cost $c_i$ within a square belonging to region $i$ can be easily computed as:
$$c_i(r_i) = 4 \lambda_i \frac{r_i^{\gamma+2}}{\gamma+2} = \zeta \lambda_i k_{i,1}^{-\frac{\gamma+2}{2}} $$
where $\zeta \triangleq 2^{(2-\gamma)/2}/(\gamma+2)$. 
Hence the total approximation cost in the whole domain, which depends on the vector $\bm k$ of cache slots $k_{i,1}$'s,
is $\expC(\bm k) = \sum_{i=1}^{M} k_{i,1} c_{i,1}(k_{i,1})$.
 
We select the values $\bm k$ that minimize the expected cost:
\begin{equation}\label{eq:opt1}
\begin{aligned}
& \underset{k_{1,1}, \dots, k_{M,1}}{\text{minimize}} & & \zeta \sum_{i=1}^{M} \lambda_i k_{i,1}^{-\gamma/2} \\
& \text{subject to} & & \sum_{i=1}^{M} k_{i,1} = k_{{1}}
\end{aligned}
\end{equation}

Employing the standard Lagrange method, one obtains that $\lambda_i k_{i,1}^{-(\gamma+2)/2}$ equals some unique constant for any region $i$, which means  
that $k_{i,1}$ has to be proportional to $\lambda_i^{2/(\gamma+2)}$. 
After some algebra we get:
\begin{equation}\label{eq:approx1}
\min \expC({\bm k}) = \zeta k_{1}^{-\gamma/2} \left( \sum_{i=1}^M \lambda_i^{\frac{2}{\gamma+2}} \right)^{\frac{\gamma+2}{2}}.
\end{equation}
In the limit of large $M$, we substitute the sum in \equaref{approx1} with an integral, obtaining:     
\begin{equation}\label{eq:approx2}
\min \expC({\bm k}) \approx \zeta k_1^{-\gamma/2} \left(\int_{\sX} \lambda(x)^{\frac{2}{\gamma+2}} \diff x \right)^{\frac{\gamma+2}{2}}.  
\end{equation}

We observe that, when the distance is the norm-1, this approach from~\cite{infocom2020} can be extended to higher dimensions  computing integrals similar to~\eqref{eq:cr}.\footnote{In the $d$ dimensional case we 
have $c(r)= a_d \lambda r^{\gamma+d}$, for an appropriate constant $a_d$.}
Under other distances, things are not as simple, but in principle one can
determine the best partitioning of the domain into $k_1$ Voronoi cells\footnote{This task
is not hard when the domain $\sX$ can be exactly partitioned into $k_1$ Voronoi cells of the same shape.
Otherwise, for sufficiently large cache sizes, one can neglect border effects and 
approximately consider $k_1$ Voronoi cells of the same shape covering the entire domain.} $V_i$  
with center $b_i$, such that $$ \expC(\mathcal A) = \sum_i \int_{V_i} C_a(x,b_i) \diff x $$ is minimum, 
and store in the cache objects $\{b_i\}_i$.   
Similarly to \cite{infocom2020}, we prefer to avoid such geometric complications, and stick for simplicity 
to the norm-1 case. 

\subsection{Chain topology}
Here we extend the approach recalled in previous section to a chain network of $N$ caches, where 
requests arrive at the leaf cache 1, and are possibly forwarded along the chain up to the node providing the best approximizer.
In a chain the cost incurred by request $r$ for object $x$, served by approximizer $\alpha = (o',j)$ 
is $C(r,\alpha) = C_a(x,o') + h(1,j)$. As request originates always at the leaf cache $1$, we simplify the notation and denote $h(1,j)$  by $h_j$.
We naturally assume $h_i > h_j$ if $i > j$.  
The $N$-th cache in the chain is the repository, where the approximation cost is negligible. In the following formulas, we recover this situation considering that the last cache has infinite cache size.

Let $k_{i,j}$ be the number of cache slots devoted by cache $j$ to region $i$.
Each of these slots is used to approximate requests falling in a square 
of area $1/k_{i,j}$ and radius  $r_{i,j} = \sqrt{1/(2 k_{i,j})}$.
Hence the cost incurred by requests falling in a square of region $i$ and served by cache $j$ is:
\begin{multline}\label{eq:cij}
c_{i,j}(r_{i,j}) = 4 \int_{0}^{r_{i,j}}\int_{0}^{r_{i,j}-x} [(x+y)^\gamma +h_j] \lambda_i \diff y \diff x  = \\ 
4 \lambda_i  \frac{r_{i,j}^{\gamma+2}}{\gamma+2} + 2 \lambda_i r_{i,j}^2 h_j
\end{multline}

The cost $C_{i,j}$ incurred by all requests falling in region $i$ and served by cache $j$, as function of $k_{i,j}$, reads:
\begin{equation}\label{eq:Cij}
C_{i,j}(k_{i,j}) = \zeta \lambda_i k_{i,j}^{-\frac{\gamma}{2}} + \lambda_i h_j
\end{equation}

{\color{black}  In general a region $i$ can be served by several caches along the path (every cache for which  $k_{i,j}>0$).  However observe that  a single request (i.e., a point of the region) will be always served by one specific  cache, cache $j^*$  with $j^* = \arg\!\min_j C_{i,j}$  (ties can be neglected). 
We encode previous property by introducing weights $w_{i,j} \in [0,1]$, where 
$w_{i,j}$ represents the fraction of region $i$ served exclusively by cache $j$. 
Let ${\bm w}_j$ be the vector of $\{w_{i,j}\}_i$.}


We obtain the optimization problem:
\begin{equation}
\begin{aligned}\label{eq:optZrelax}
& \underset{{\bm w}_2, \dots, {\bm w}_N}{\text{minimize}} & &  
\zeta k_1^{-\gamma/2} \Bigg(\sum_{i=1}^M \bigg(1-\sum_{j=2}^N w_{i,j}\bigg) \lambda_i^{\frac{2}{\gamma+2}} \Bigg)^{\frac{\gamma+2}{2}} + \\
&&& \sum_{i=1}^M \bigg(1-\sum_{j=2}^Z w_{i,j}\bigg) w_i \lambda_i h_1  + \\
&&& \sum_{j = 2}^N \Bigg[ \zeta k_j^{-\gamma/2} \bigg(\sum_{i=1}^M w_{i,j} \lambda_i^{\frac{2}{\gamma+2}} \bigg)^{\frac{\gamma+2}{2}} + \sum_{i=1}^M w_{i,j} \lambda_i h_j \Bigg] \\
& \text{subject to} & & w_{i,j} \geq 0 \qquad \forall j > 1, \forall i \\
&&& \sum_{j=2}^N w_{i,j} \leq 1 \qquad \forall i 
\end{aligned}
\end{equation}
where notice that we have separated the contribution of cache 1, and taken as decision
variables vectors ${\bm w}_j$, with $j > 1$, since 
${\bm w}_1 = {\bm 1} - \sum_{j=2}^N {\bm w}_j$.
Moreover, notice that the constraints in \equaref{optZrelax} are sufficient to guarantee
that also the following obvious constraints hold:
\begin{eqnarray*}
w_{i,j} \leq 1  && \quad \forall j > 1, \forall i \\
0 \leq w_{i,1} \leq 1 && \quad \forall i
\end{eqnarray*} 
In this form, \equaref{optZrelax} is a convex
minimization problem over a convex domain, thus it has a global minimum.   
Without loss of generality, let the $M$ regions be sorted in increasing values of $\lambda_i$.
Employing the standard method of Lagrange multipliers, 
KKT conditions imply that the global optimum is attained when
cache 1 handles all most popular regions region $i > i^*$ (i.e., $w_{i,1} = 1$, $i>i^*$), plus possibly a 
piece of region $i^*$ (if $0 < w_{i^*,1} < 1$). Cache 1 does not allocate any slot to regions $i < i^*$.

Previous result allows us to prove the following interesting property about the structure of the 
optimal solution:  

\begin{prop}\label{prop-chain} In the case of a chain topology,  with requests arriving only
at the first cache, the best solution of the continuous-domain, finite-$M$ problem 
\equaref{optZrelax} is characterized by a set of popularity thresholds $\lambda_0^* = \min\{\lambda_i\} \leq \lambda_1^* \leq  
\lambda_2^* \leq \ldots \leq \lambda_{N-1}^* \leq \lambda_N = \max\{\lambda_i\} $, such that cache $j$ approximates all requests
falling in regions $i$ with $\lambda_{j-1}^* < \lambda_i < \lambda_{j}^*$, plus possibly
a portion of a region with $\lambda_i = \lambda_{j-1}^*$, and a portion of 
a region with $\lambda_i = \lambda_{j}^*$.
\end{prop}
\begin{proof}
It is sufficient to apply the above property about the regions handled by cache 1,
filtering out the requests handled by cache 1, and iteratively applying the same result
to the request process forwarded upstream to caches $2,\ldots,N$.  
\end{proof}

When the set of popularity values is not finite, it is possible to extend the result in Proposition~\ref{prop-chain}, 
letting $M$ diverge. We partition $\sX$ into $N$ sub-domains 
${\sX_j}$, $j = 1,\ldots,N$, stacked in vector $\bm \sX$, such that cache $j$ handles only requests falling into domain
$\sX_j$, and we seek to minimize:
\begin{equation}\label{eq:approx-net}
\expC({\bm \sX}) = \sum_{j=1}^N \left[\zeta k_j^{-\gamma/2} \left(\int_{\sX_j} \!\! \lambda(x)^{\frac{2}{\gamma+2}} \diff x \right)^{\frac{\gamma+2}{2}}
\!\!\!\!+ h_j \int_{\sX_j} \lambda(x) \diff x \right] \nonumber
\end{equation}
In principle we would like to find the best partitioning:   
\[
{\bm \sX}^* =\argmin_{\bm \sX}  \expC({\bm \sX})  
\]
In this asymptotic case we can restate Proposition \ref{prop-chain} as follows, providing a simpler and more elegant proof.
\begin{prop}\label{prop-chain2} In the case of a chain topology with requests arriving only
at the first cache,  the best partition ${\bm \sX}^*$ is characterized by the following property: 
for any $i<j$, $\inf_{\sX_i^*} \lambda(x) \ge \sup_{\sX_j^*}\lambda(x)$. 
\end{prop}
\begin{proof}
By contradiction, let us assume that we find two non negligible areas $\Delta \sX_i \subseteq \sX_i^*$ and  $\Delta \sX_j\subseteq \sX_j^*$
such that:
\begin{equation} 
\sup_{\Delta \sX_j}\lambda(x) >  \inf_{\Delta \sX_i}\lambda(x) \nonumber
\end{equation}
Then we can always find two non-negligible areas  $\Delta \sX_i' \subseteq \Delta \sX_i$ 
and  $\Delta \sX_j' \subseteq \Delta \sX_j$ such that we jointly have:
\begin{equation}\label{eq:commint}
\int_{\Delta \sX_i'}\!\!\lambda(x)^{\frac{2}{2+\gamma}} \diff x  = \int_{\Delta \sX_j' }\!\!\lambda(x)^{\frac{2}{2+\gamma}} \diff x 
\end{equation}
and
\begin{equation} \label{eq:sepa}
\inf_{\Delta \sX_j'}\lambda(x) \ge  \sup_{\Delta \sX_i'}\lambda(x) > 0 
\end{equation}

Now let us see what happens if we \lq swap' $\Delta \sX_i'$ with $\Delta \sX_j'$, i.e., 
if we take a new partition ${\bm \sX'}$ where $\sX_i'=(\sX_i^* \setminus \Delta \sX_i') \cup  \Delta \sX_j'$ and 
$\sX_j'=(\sX_j^*\setminus \Delta \sX_j') \cup  \Delta \sX_i'$.  
Note that by construction
\begin{equation}
\expC({\bm \sX'}) = \expC({\bm \sX}^*) + (h_j-h_i) \int_{\Delta \sX_i'} \!\!\lambda(x) \diff x + (h_i-h_j) \int_{\Delta \sX_j'} \!\!\lambda(x) \diff x \nonumber
\end{equation}
Therefore, since $h_j > h_i$, we have $\expC({\bm \sX'}) \le \expC({\bm \sX}^*)$ if we can show that 
\begin{equation}
\int_{\Delta \sX_j'} \lambda(x) \diff x \ge \int_{\Delta \sX_i'} \lambda(x) \diff x. \nonumber
\end{equation}
%

Denoted with $\beta= 2/(2+\gamma)<1$ we have:
\begin{align*}
 & \int_{\Delta \sX_j'} \lambda(x) \diff x =  \int_{\Delta \sX_j'} \lambda(x)^{\beta} \lambda(x)^{1-\beta} \diff x & \\
 & \ge (\inf_{\Delta \sX_j'} \lambda(x))^{1-\beta} \int_{\Delta \sX_j'} \lambda(x)^{\beta}   \diff x & \\
 &= (\inf_{\Delta \sX_j'} \lambda(x))^{1-\beta}  \int_{\Delta \sX_i'} \lambda(x)^{\beta}   \diff x  \qquad \qquad & \textrm{by \equaref{commint}} \\
 & \ge (\sup_{\Delta \sX_i'} \lambda(x))^{1-\beta}  \int_{\Delta \sX_i'} \lambda(x)^{\beta}   \diff x \qquad \qquad & \textrm{by \equaref{sepa}}  \\
 & =  \int_{\Delta \sX_i'} (\sup_{\Delta \sX_i'} \lambda(x))^{1-\beta} \lambda(x)^{\beta}   \diff x & \\
 & \ge \int_{\Delta \sX_i'}  \lambda(x)^{1-\beta} \lambda(x)^{\beta}   \diff x = \int_{\Delta \sX_i'}  \lambda(x)   \diff x &
\end{align*}
\end{proof}

\subsection{Extension to equi-depth trees}
Previous results obtained for the chain topology can be easily extended
to trees with $L$ leaves at the same depth $D$, where requests arrive only at the leaves
and all caches at the same level have the same size.
Let $h_{D-j}$ be the (equal) cost to reach the cache at level $j$ starting from a leaf.
{\color{black} We assume the spatial arrival rate at leaf $\ell$ to be given by  $\lambda_\ell(x)= \beta_\ell \lambda(x)$,  for some constant $\beta_\ell>0$, i.e., spatial arrival rates at different caches are identical after rescaling by a constant factor.  Moreover arrival processes at different leaves are assumed to be independent.}
We will call {\em equi-depth} tree a cache network with the above characteristics. 
We naturally assume $h_i > h_j$ if $i > j$. 
    
\begin{prop}
In an equi-depth tree the optimal cost is achieved
by replicating the same allocation at each cache of the same level.
The allocation to be replicated is the one obtained in the special
case of a chain topology ($L = 1$).
\end{prop}

\begin{proof}
Suppose to increase the number of nodes in the topology, creating a system of 
$L$ parallel chain topologies. Each leaf now has an independent path towards
a dedicated copy of the root node. By doing so the total cost in the system of parallel chains
is surely not larger than the total cost achievable in the original tree, and, in general,
it might be smaller (this because we can independently place objects in every chain  
so as to minimize the cost induced by the requests arriving at the corresponding leaf).
On the other hand, the optimal allocation on each chain is the same, since 
the objective function in \equaref{optZrelax} is linear with respect to parameter $\beta_\ell$.
Therefore, by adopting such equal allocation on each cache of the same level in the original
tree, we obtain exactly the same total cost achieved in the system of parallel chains, hence
this allocation is optimal.  
\end{proof}

\subsection{A tandem network with arrivals at both nodes}
In general cache networks that do not belong to the class of equi-depth trees, the 
simple optimal structure described in Proposition \ref{prop-chain} is, unfortunately, lost.
To see why, it is sufficient to consider the simple case of a tandem network with 
two identical caches (hereinafter called the leaf and the parent), where the same
external arrival process $\lambda(x)$ of requests arrives at both nodes.
Now, let us suppose that the cost $h$ to reach the parent from the leaf is large (but it does not need to be 
disproportionally large). Then the leaf will not find particularly convenient to forward its requests 
to the parent, unless maybe for objects very close to the ones stored in the parent (whichever they are).
On the other hand, the parent has to locally approximate all requests, hence it will need to adequately cover
the entire domain $\sX$ like an isolated cache. As a consequence, we do not expect any clear separation
of $\sX$ into a sub-domain handled by the leaf, and a sub-domain handled by the parent.
In particular, the property that we had before, according to which a single cache has to allocate slots
to cover a particular region of the domain, does not hold anymore.

A more formal explanation of what happens in this simple case 
can be provided by the following model. 
Again, we divide the domain, both at the leaf and at the parent cache, into $M$ regions of unitary area. 
The request rate over each region is assumed to be constant and we denote it by $\lambda_i$ and $\beta \lambda_i$ for the leaf 
and the parent cache, respectively (hence by setting $\beta = 0$ we can recover previous case in which requests arrive only at the leaf).
Let $k_{i,1}$ and $k_{i,2}$ be the number of slots devoted to region $i$ by the leaf and the parent node, respectively.
Notice that now both quantities are in general different from zero. 
The leaf node will forward to the parent the requests falling in a fraction $(1-w_{i,1})$ of region $i$, and it is natural
to assume that these requests are those falling farther from the locally stored objects, i.e., 
at a distance larger than $r^*_{1,i} = \sqrt{w_{1,i}} r_{1,i}$, where $r_{1,i} = \sqrt{1/(2 k_{i,1})}$. 
Therefore the approximation cost \equaref{Cij} is changed to:
$$C_{i,1}(k_{i,1},w_{i,1}) = \zeta \lambda_i {w_{i,1}}^{\frac{\gamma+2}{2}} {k_{i,1}}^{-\frac{\gamma}{2}}. $$   
 
 

Requests forwarded to the parent cache will experience an additional movement cost $h$, 
plus a local approximation cost at the parent, that we model by assuming that
the total area of the subregion forwarded to the parent cache  $k_{i,1} 2 r_{1,i}^2 (1-{w_{1,i}})$ will be served 
by the $k_{i,2}$ points at the parent, within squares of radius: 
\[\sqrt{ \frac{k_{i,1} r_{i,1}^2 (1-{w_{i,1}})}{k_{i,2}}} = \sqrt{\frac{1-{w_{i,1}}}{ 2 k_{i,2}}}\]  
Moreover, at the parent cache the local requests will generate an approximation cost similar to \equaref{Cij} (with no 
retrieval cost). 

The total approximation cost in the network is then:
\begin{multline*}
\expC(A) = \zeta \sum_{i=1}^M \lambda_i {w_{i,1}}^{\frac{\gamma+2}{2}} {k_{i,1}}^{-\frac{\gamma}{2}} \\
+  \zeta \sum_{i=1}^M \lambda_i (\beta + (1-w_{i,1})^{\frac{\gamma +2}{2}}) {k_{i,2}}^{-\frac{\gamma}{2}} + h \sum_{i=1}^M \lambda_i (1- w_{i,1}).
\end{multline*} 
This cost should be minimized over $\{w_{i,1}\}_i$, $\{k_{i,1}\}_i$, and $\{k_{i,2}\}_i$.
By finding the optimal values for $\{k_{i,1}\}_i$ and $\{k_{i,2}\}_i$ given $\{w_{i,1}\}_i$, we get
\begin{align}\label{eq:tandemasym}
\expC({\bm  w}) & = \zeta k_1^{-\frac{\gamma}{2}} \left(\sum_{i=1}^M \lambda_i^{\frac{2}{2+\gamma}} {w_{i,1}}\right)^{\frac{2+\gamma}{2}} \nonumber\\
	& + \zeta k_2^{-\frac{\gamma}{2}} \left(\sum_{i=1}^M \lambda_i^{\frac{2}{2+\gamma}} (\beta + (1-w_{i,1})^{\frac{\gamma + 2}{2}})^{\frac{2}{2+\gamma}} \right)^{\frac{2+\gamma}{2}} \nonumber\\
	& + h \sum_{i=1}^M \lambda_i (1- w_{i,1}).
\end{align} 
Note that for $\beta = 0$ we recover the cost resulting from \equaref{optZrelax} in the case of a tandem network. 
Computing the derivative of the above cost with respect to $w_{i,1}$ we get:
\begin{align}
& \frac{\partial \expC( w)}{\partial w_{i,1}} 
= \zeta k_1^{-\frac{\gamma}{2}} \frac{\gamma+2}{2} \left(\sum_{i=1}^M \lambda_i^{\frac{2}{2+\gamma}} {w_{1,i}}\right)^{\frac{\gamma}{2}}   \lambda_j^{\frac{2}{2+\gamma}} \nonumber\\
	& - \zeta k_2^{-\frac{\gamma}{2}}  \frac{\gamma+2}{2}\left(\sum_{i=1}^M \lambda_i^{\frac{2}{2+\gamma}} (\beta + (1-w_{1,i})^{\frac{\gamma + 2}{2}})^{\frac{2}{2+\gamma}} \right)^{\frac{\gamma}{2}} \nonumber\\
	& \times \lambda_j^{\frac{2}{2+\gamma}} \frac{(1 - w_{1,j})^{\frac{\gamma}{2}}}{(\beta + (1-w_{1,j})^{\frac{\gamma + 2}{2}})^{\frac{\gamma}{2+\gamma}}}
	- h \lambda_j.
\end{align} 
Imposing the optimality conditions, we find that there may be multiple regions with different popularities $\lambda_i$ for which $w_{1,i}* \in (0,1)$, i.e., for which the leaf forwards part of the requests to the parent. The
structure of the solution in Proposition \ref{prop-chain} might be lost, leading
to optimal allocations where both caches handle portions of the same region.

\begin{figure}
\centering
\includegraphics[scale=0.6]{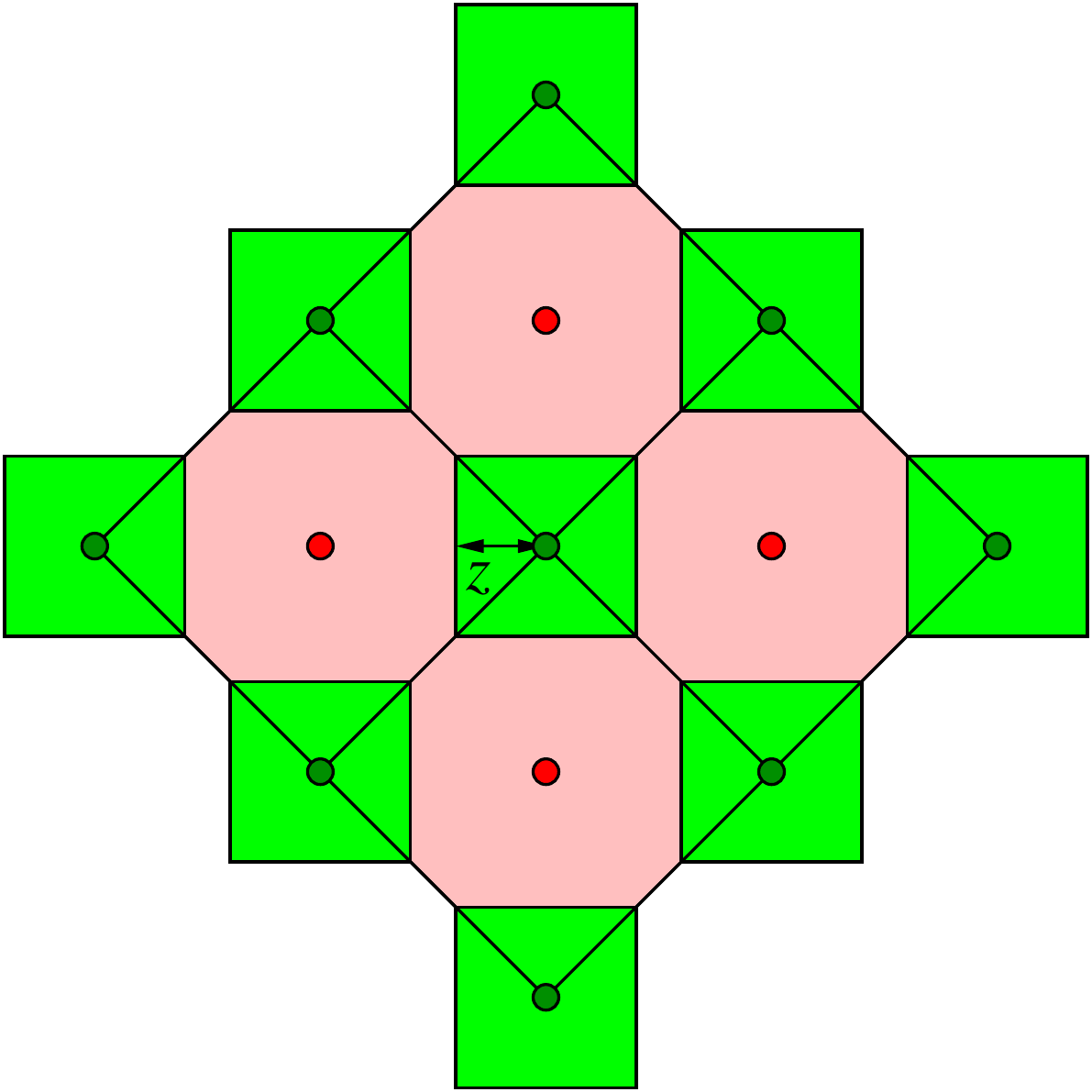}
\caption{Optimal allocation in the tandem network with uniform arrival process at both nodes: 
square tessellation in the leaf (red nodes), and square tessellation in the parent (green nodes).}
\label{fig:piastrelle}
\end{figure}

To shed light into this phenomenon, we have further investigated the special
case in which $\lambda$ is uniform over the whole domain. In this case it is convenient to shift over space the two regular tessellations so that the centroids at the leaf and at the parent are as far as possible, as shown in Fig.~\ref{fig:piastrelle}.
This allows the leaf to forward the requests farthest from its centroids to the parent, where they are better approximated.


Requests arriving at the leaf are approximated by the leaf in the red 
portion of the domain, as depicted in Fig. \ref{fig:piastrelle}, while 
they are approximated by the parent in the green portion of the domain.
Distance $z$ (in Fig. \ref{fig:piastrelle}) that defines the separation between the two portions 
can be easily computed (for $\gamma = 1$) as $z = \max\{0,(r-h)/2\}$, where $r$ is the radius
of the square of each tessellation (note that if $h > r$ requests
are not forwarded from the leaf to the parent). Then one can easily compute
the reduction $\Delta c = \frac{8}{3}z^3$ in the approximation cost
for requests arriving at the leaf, provided by each slot of the second cache,
and compute the resulting overall approximation cost (the approach
can be generalized to $\gamma \neq 1$, but we omit the details here).

\section{\netduel: an online dynamic policy}\label{sec:netduel}
Although in our work we have focused on the static, offline problem of content 
allocation at similarity caches, we have also devised an online, $\lambda$-unaware dynamic
policy 
\netduel{}, which is a networked version of policy \duel{} \textcolor{black}{we have} proposed in \cite{infocom2020}. 
At high-level, it is based on the 
following idea: each (real) content currently is the cache is paired to a (virtual\footnote{The cache stores only metadata of a virtual object, not the object itself. Virtual objects are taken from the arrival process.})
content competing with it. The cumulative saving in the total cost 
produced by the real and the virtual objects are observed over a suitable time window,
and if the saving of the virtual object exceeds the saving of the real one by a sufficient amount,
the virtual replaces the real in the cache. 
{\color{black} Otherwise, at the end of the observation window, the virtual object is discarded,
and afterwards the real object will be paired to a new virtual object taken from the arrival process.} 
\netduel{} achieves an allocation close to the optimal one, suggesting that
effective online dynamic policies can be devised for networks of similarity caches,
at least under the assumption that each node knows when to forward requests 
upstream.
\section{Numerical experiments}
\subsection{Synthetic arrival process}
To test our algorithms, we consider $10000$ objects falling on the points 
of a bi-dimensional $L \times L$ grid with $L=100$, equipped with the norm-1 metric and the
local cost $C_a(x,y) = d(x,y)$, i.e., we take (unless otherwise specified) $\gamma = 1$. 
The request process follows a Gaussian distribution, such that the request
rate of object $i$ is proportional to $\exp(-d_i^2/(2 \sigma^2))$, 
where $d_i$ is the hop distance from the grid center.
To jointly test our continuous approximations, we assume that each grid point $i$
is the center of a small square of area 1, on which $\lambda$ is assumed to be constant
and equal to $\lambda_i$. 

\begin{figure}
\centering
\includegraphics[scale=0.4]{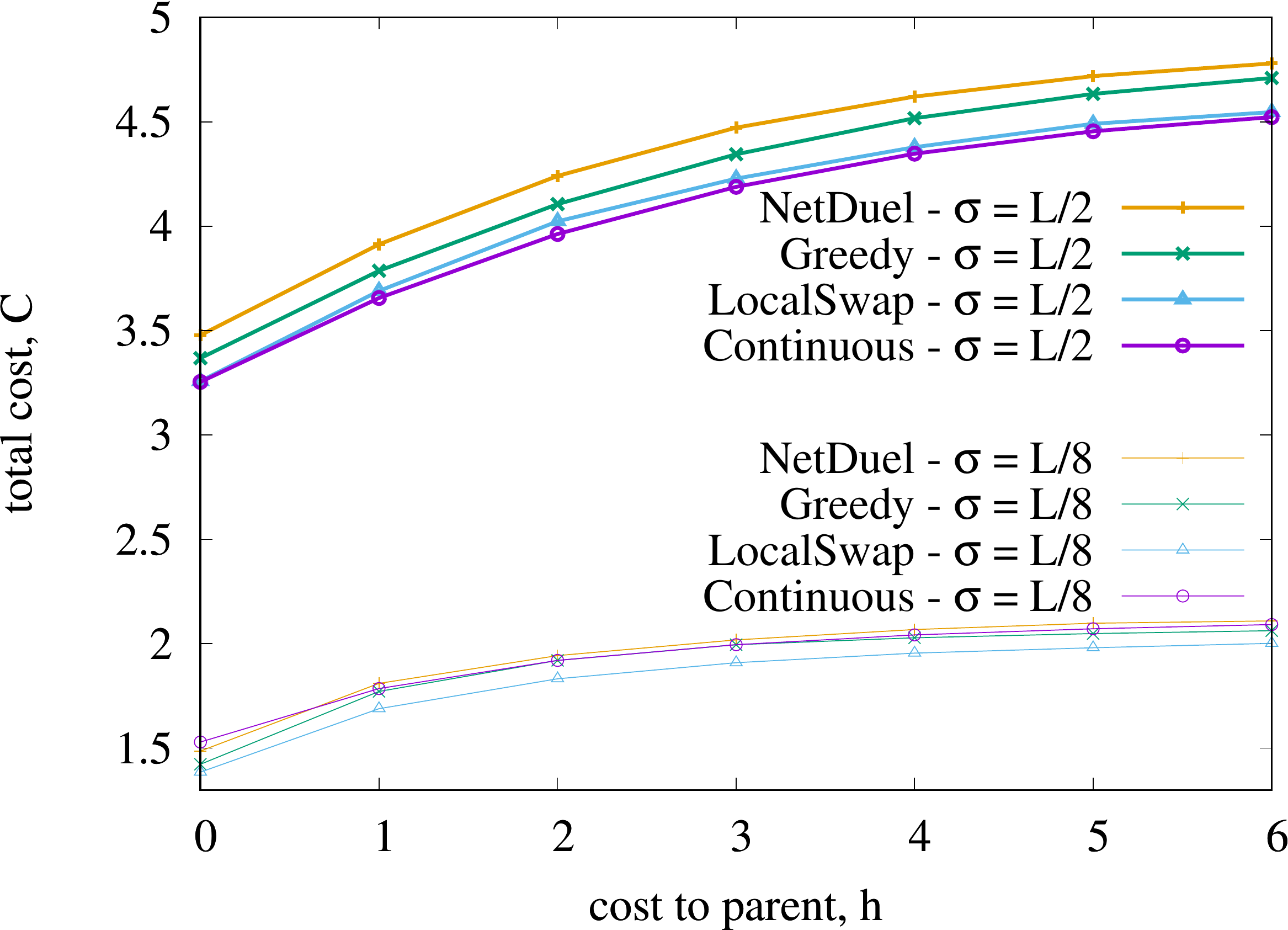}
\caption{Total cost obtained by \greedy{}, \localswap{}, continuous approximation and \netduel{} in a tandem network
with arrivals at the leaf, for $\sigma = L/2$ (thick curves) or $\sigma = L/8$ (thin curves).}
\label{fig:fig2}
\end{figure}

We first consider a simple tandem network with arrivals only at the leaf, and fixed cost
$h$ to reach the parent. 
In Fig. \ref{fig:fig2} we compare the total cost produced by \greedy, \localswap, the continuous approximation
(the solution of \equaref{optZrelax}) and \netduel, as function of $h$, for a larger gaussian ($\sigma = L/2$)
or a narrow gaussian ($\sigma = L/8$). We observe that \localswap{} performs better than \greedy{}, which performs better
than \netduel. The continuous approximation does not necessarily provide a lower bound to discrete algorithms/policy,
since it is a different system where the request space is continuous, rather than constrained on the grid points.
However, we do observe that the continuous approximation curve gets closer to the curve produced by \localswap{}
for $\sigma = L/2$ (thick curves), since in this case $\lambda$
varies more smoothly over the domain.   

\begin{figure}
\centering
\includegraphics[scale=0.22]{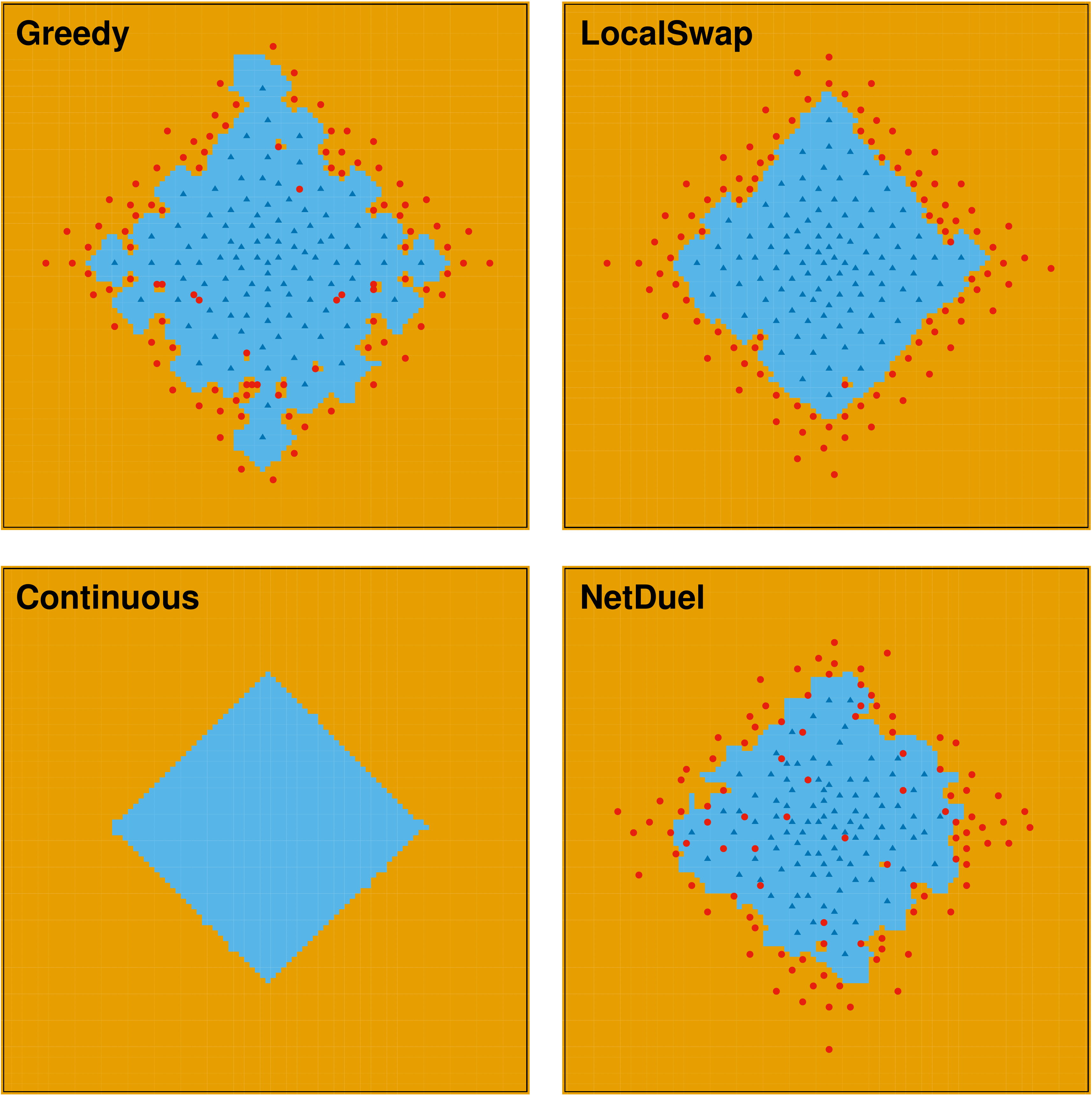}
\caption{Allocations obtained by \greedy{}, \localswap{}, continuous approximation and \netduel{} in the tandem network
with $\sigma = L/8$, $h=3$. Circle marks for the parent cache and triangle marks for the leaf cache. }
\label{fig:topo}
\vspace{-2mm}
\end{figure} 

In Fig. \ref{fig:topo} we show the allocations (circles for the parent, triangles for the leaf) 
produced by the four approaches above in the case $\sigma = L/8$ and $h=3$, using two different colors for the 
sub-domains where requests arriving at the leaf are approximated by the leaf or the parent\footnote{For the
continuous approximation, we do not show stored contents, and (border) squarelets are considered as handled
exclusively by the parent if $w_{i,2} > w_{i,1}$.}.
We observe that \greedy{} and \netduel{} produce more irregularities than \localswap{}, as compared
to the theoretical prediction of the continuous approximation.
 
\begin{figure}
\centering
\includegraphics[scale=0.25]{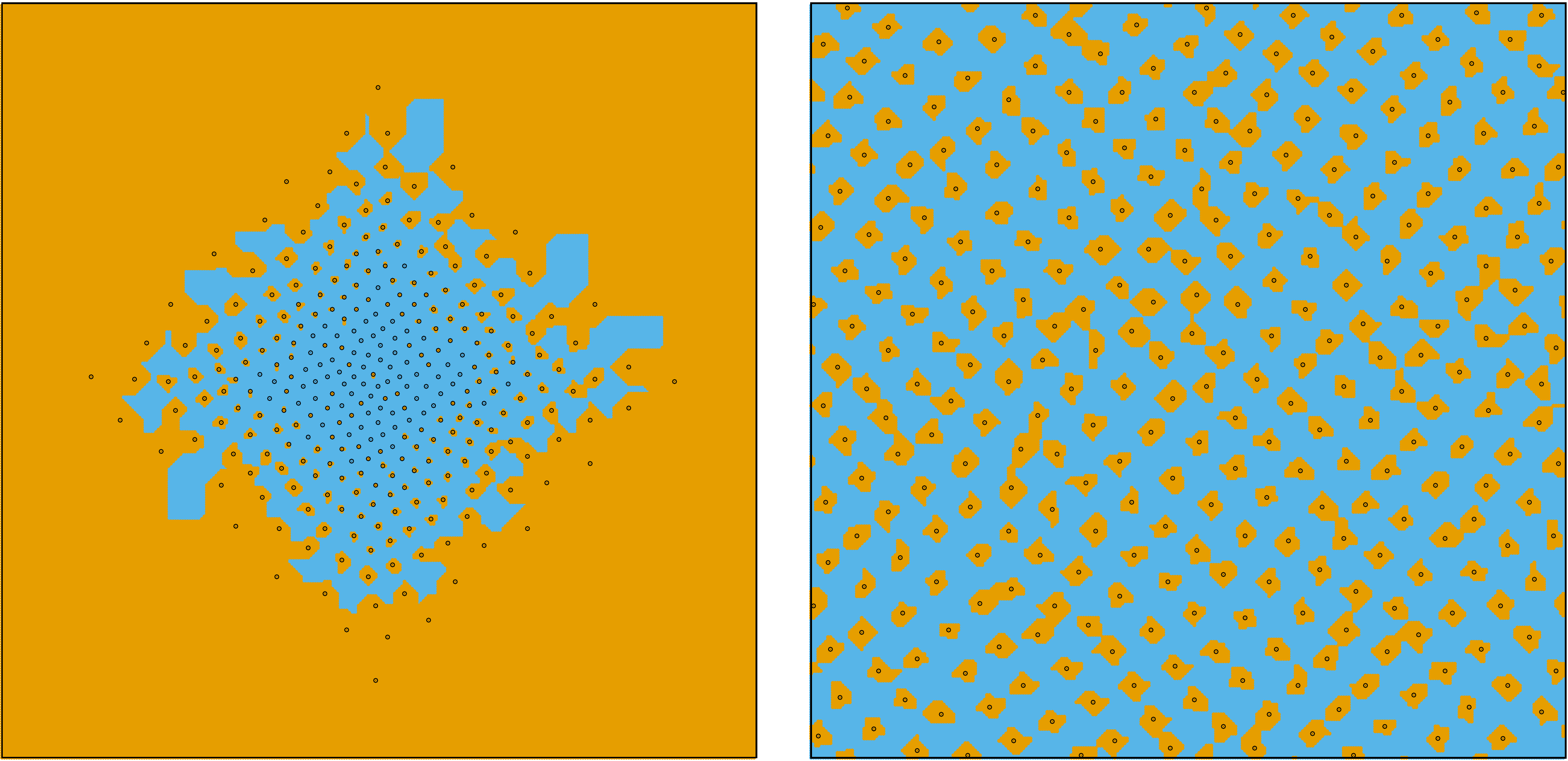}
\caption{Parent allocation obtained by \localswap{} in a tandem network with arrivals at both nodes.
Gaussian traffic (left plot) and Uniform traffic (right plot).}
\label{fig:topo2}
\vspace{-2mm}
\end{figure}

\begin{figure}
\centering
\includegraphics[scale=0.4]{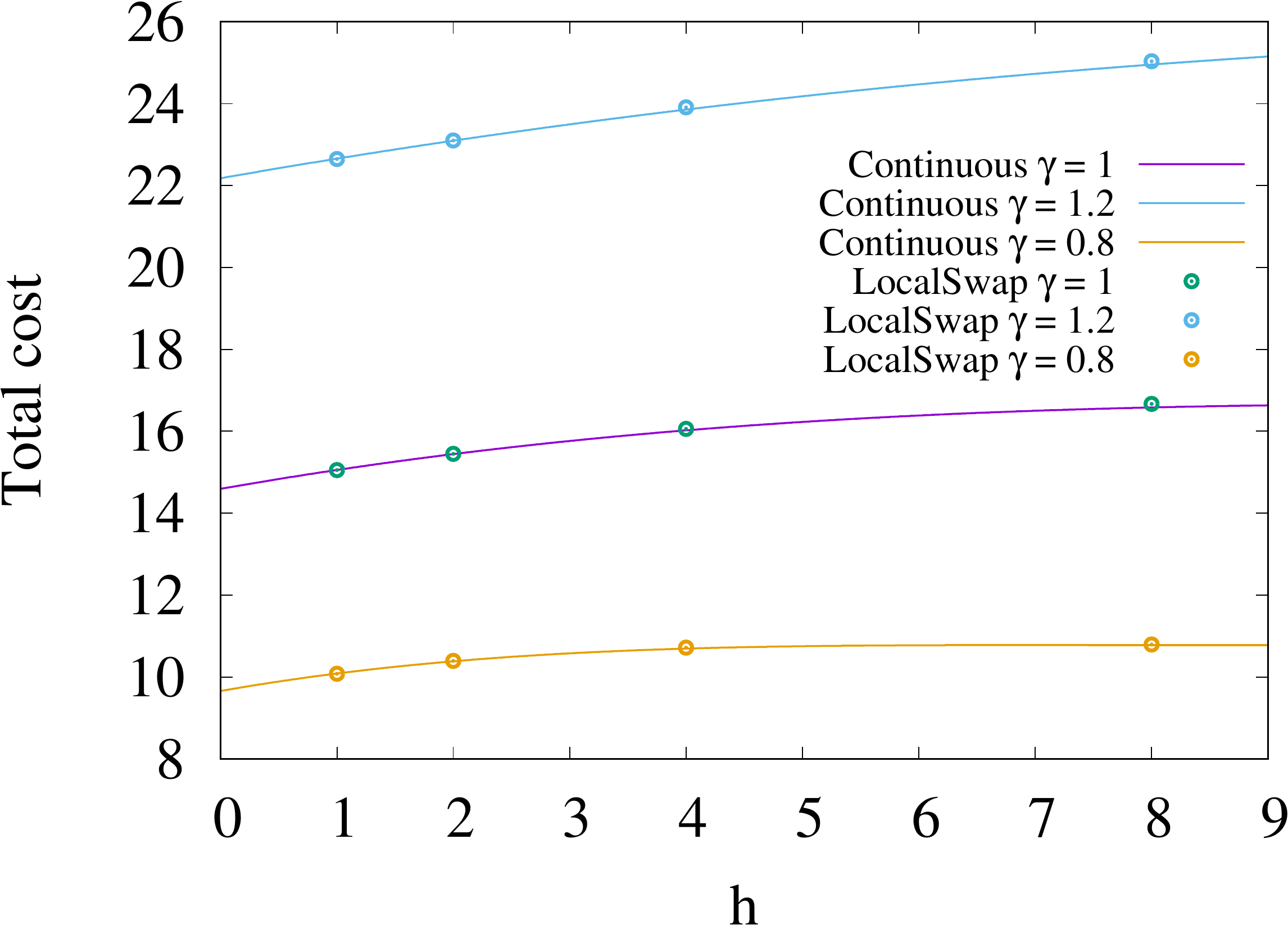}
\caption{Total cost in the tandem network with arrivals at both node, $\lambda$ uniform, as
function of $h$, for different values of $\gamma$, according to \localswap{} (points) and continuous approximation (curves).}
\label{fig:plot1}
\end{figure}

In Fig. \ref{fig:topo2} we report, for a larger system with $100000$ contents, the allocation produced 
at the parent by \localswap{} in a tandem network with requests arriving at both nodes, showing 
also with two different colors the regions where requests arriving at the leaf are approximated by the leaf or the parent.
We consider both a Gaussian arrival process with $\sigma = L/8$ (left plot), and a simple 
Uniform process (right plot), and fixed $h = 3$. Notice that the parent cache covers also the central part
of the domain, in contrast to Fig. \ref{fig:topo}.
Results produced by \localswap{} suggest that now,
for the requests arriving at the leaf, the regions served directly by the leaf and the regions approximated 
by the parent are intertwined in a complex way.
For uniform $\lambda$, Fig. \ref{fig:plot1} shows the accuracy of the continuous approximation
based on the shifted regular square tessellations shown in Fig. \ref{fig:piastrelle}.    
  
\subsection{Amazon trace}\label{subsec:amazon}
By crawling the Amazon web-store, the authors of \cite{McAuley_amazon} built an image-based 
dataset of users’ preferences for millions of items.
Using a neural network pre-trained on ImageNet, each item is embedded into a 
$d$-dimensional space, on which Euclidean distance is used as item similarity.
We consider as request process the timestamped reviews left by users for the 10000
most popular items belonging to the baby category, with $d=100$. 
The resulting trace, containing about 10.3M requests, is fed into a 
cache of size $100$, with a parent cache of the same size (a tandem network)
reachable by paying an additional fixed cost $h=150$. The local approximation cost
is set equal to the Euclidean distance. 

\begin{figure}
\centering
\includegraphics[scale=0.32]{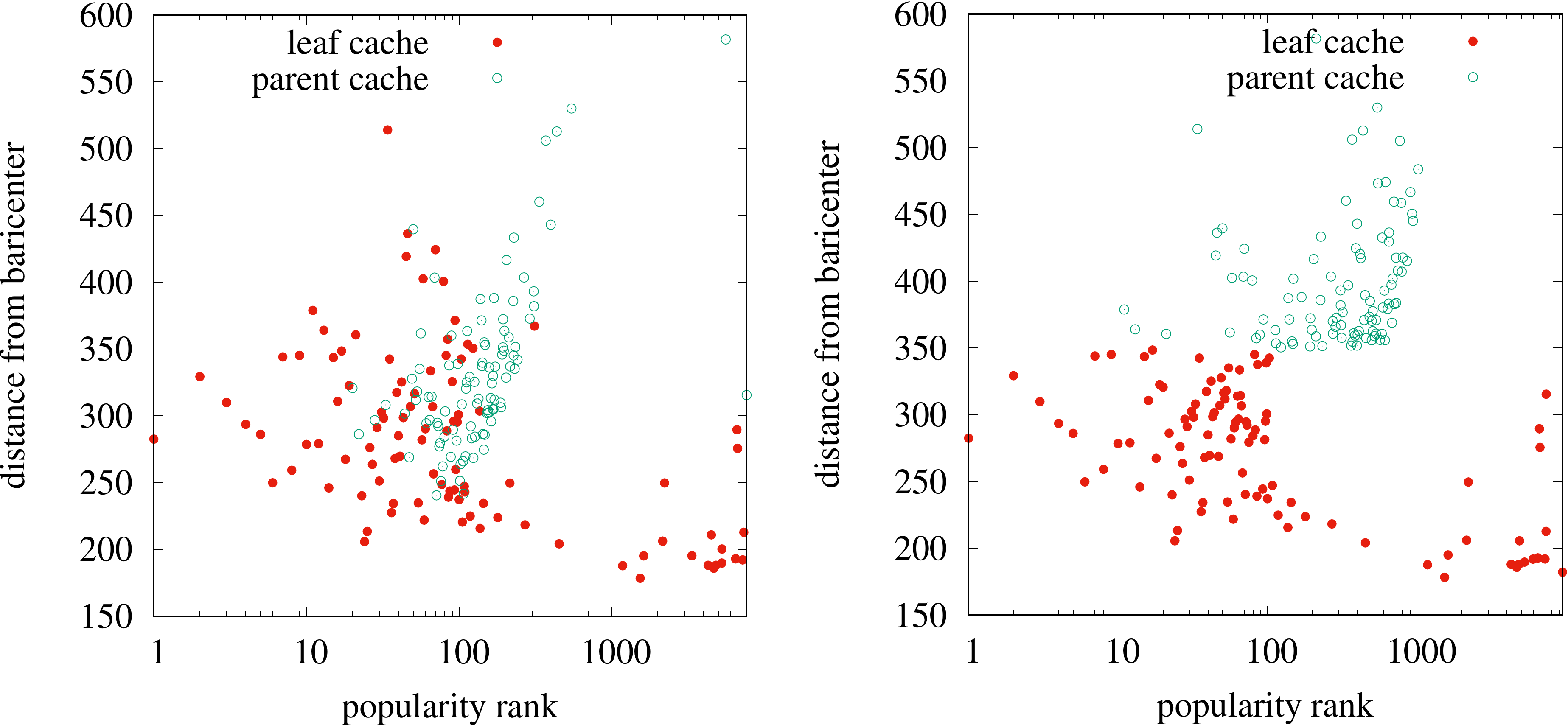}
\caption{Allocations obtained by \localswap{} in a tandem network with arrivals at the leaf 
according to Amazon trace. Unconstrained version (left) and constrained version (right).}
\label{fig:summary}
\vspace{-1mm}
\end{figure}

\begin{figure}
\centering
\includegraphics[scale=0.4]{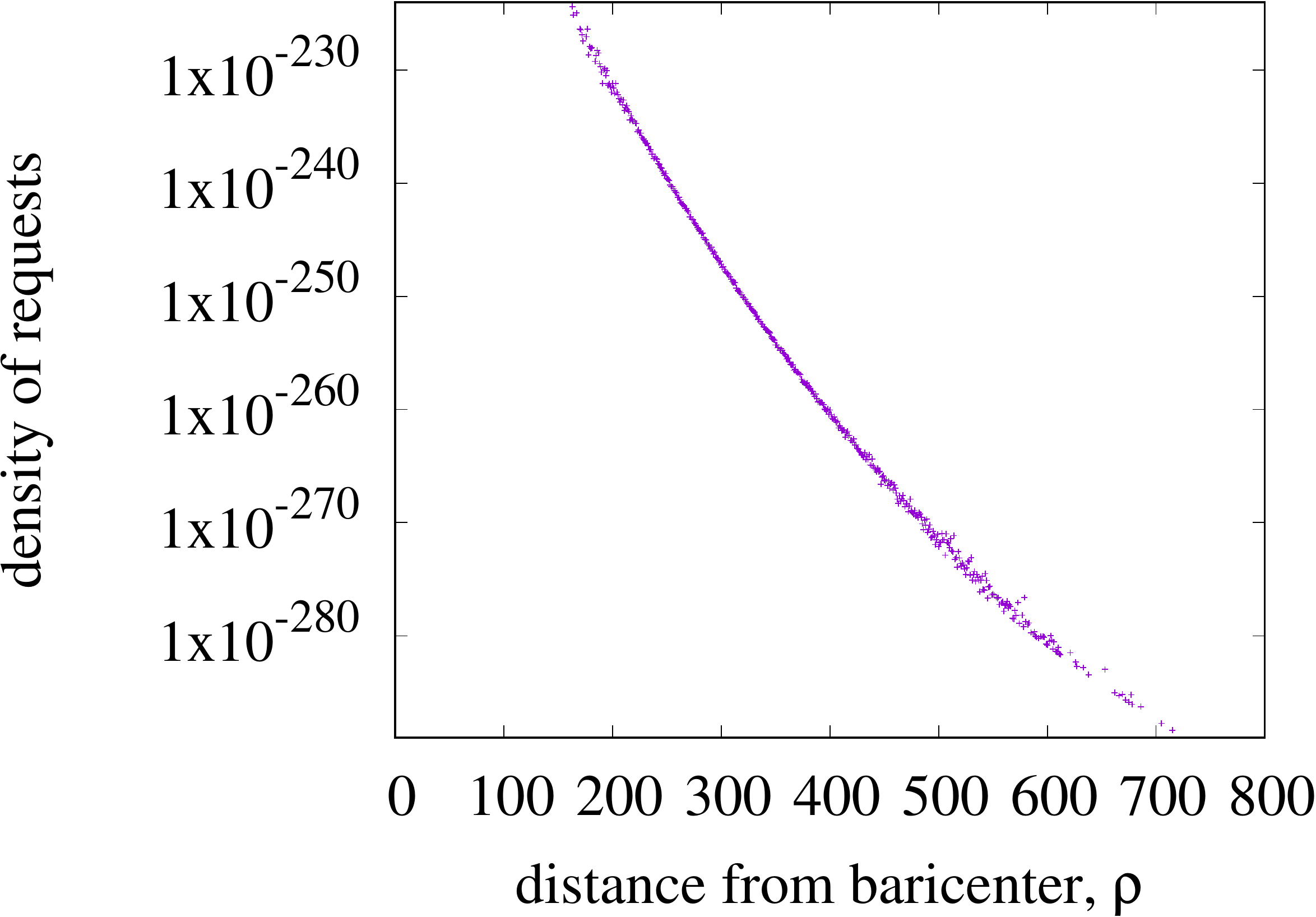}
\caption{Density of requests of Amazon trace within spherical shells at distance $d \in [\rho_k,\rho_{k+1}]$ 
from the baricenter.}
\label{fig:reqdens}
\vspace{-1mm}
\end{figure}

In Fig. \ref{fig:summary} we show the allocations produced by \localswap{} in both caches,
reporting, for each stored item, the popularity rank ($x$ axes) and the distance
from the baricenter ($y$ axes). Across the entire catalog we found no correlation between  
popularity rank and distance from the baricenter.
Nevertheless, we do observe that the leaf cache
tends to store items that are either very popular or very close to the baricenter.
The resulting total cost is $C = 266$ (left plot in Fig. \ref{fig:summary}).

Moreover, by computing the request density
within spherical shells at distance $d \in [\rho,\rho+1]$ from the baricenter, we found
a decreasing trend in $\rho$, see Fig. \ref{fig:reqdens}, which justifies the attempt of
 \lq enforcing' the structure of the optimal solution
that we found in chain topologies fed only from the leaf. We do so
by constraining the leaf (parent) cache to
store only contents at distance from the baricenter smaller (larger) than a given threshold  $d^*$. 
The constrained \localswap{} algorithm obtains, for the best possible $d^* = 350$, 
a total cost $C = 269$ (only $1\%$ worse than before), right plot in Fig. \ref{fig:summary}, 
suggesting that a simple allocation and forwarding rule based on the distance from the baricenter 
is close to optimal also in a realistic scenario.


\vspace{-2mm}
\section{Conclusions}
We made a first step into the analysis of networks of similarity caches, focusing
on the offline problem of static content allocation. Despite the NP-hardness of the
problem, effective greedy algorithms can be devised with guaranteed performance, 
but they become prohibitive as the system size increases.
For very large request space/catalog size,  we relaxed the problem to the continuous, obtaining 
for equi-depth tree topologies an easily implementable solution with a simple structure,
which greatly simplifies the related request forwarding problem.
The above simple structure is unfortunately lost in more general networks. We have also proposed a first online
dynamic policy, though much more can be done in the design of practical online policies and
request forwarding strategies for similarity caching networks.

\appendix

\section{Proof of Proposition~\ref{p:submodularity}} 
\label{a:proof}
\begin{proof}
We first show that constraints are matroid ones. The empty set obviously belongs to $\mathcal I$, and if $\sA \subset \sB$ with $\sB \in \mathcal I$, 
then $\sA \in \mathcal I$. Finally, given two allocations with $|\sA| < |\sB|$, there exists a cache $i$ that stores 
less elements under $\sA$ than under $\sB$, i.e., such that  $\sum_{o':(o',i) \in \sA} 1 < \sum_{o':(o',i) \in \sB} 1$. Then, 
there exists an object $o$ that is stored at $i$ under $\sB$, but not 
under $\sA$. As $\sum_{o':(o',i) \in \sA} 1 < \sum_{o':(o',i) \in \sB} 1 \le k_i$, $\sA\cup (o,i)$ is still a feasible allocation.

We now prove that $G(\sA)$ is a non-negative monotone submodular function. 
\begin{align*}
    G(\sA) & = \sum_r \lambda_r C(r,\emptyset) - \sum_r \lambda_r C(r,\sA) \\
        & = \sum_r \lambda_r \left(C(r,\emptyset) - C(r,\sA)\right)\\
        & = \sum_r \lambda_r \left(C(r,\emptyset) - \min_{\alpha \in S \cup \sA}C(r,\alpha)\right)\\
        & = \sum_r \lambda_r \left(C(r,\emptyset) - \min\left(\min_{\alpha \in \sA }C(r,\alpha), C(r,\emptyset)\right)\right)\\
        & = \sum_r \lambda_r \left(C(r,\emptyset) - \min_{\alpha \in \sA} \min \left( C(r,\alpha), C(r,\emptyset)\right)\right)\\
        & = \sum_r \max_{\alpha \in \sA} \lambda_r  \Big(C(r,\emptyset) -  \min \left( C(r,\alpha), C(r,\emptyset)\right)\Big)\\
        & = \sum_r \max_{\alpha \in \sA} \lambda_r  \Big( \max \left( C(r,\emptyset) - C(r,\alpha), 0 \right)\Big)
\end{align*}
Then $G(\sA) = \sum_r \max_{\alpha \in \sA} M_{r,\alpha}$, where $M_{r,\alpha}\ge 0$ for all $r$ and $\alpha$. The set function is 
obviously monotone (i.e., if $\sA \subset \sB$, then $G(\sA) \le G(\sB)$) and non-negative and corresponds to the utility of a facility location problem 
that is known to be submodular (e.g.,~\cite{krause14survey}, but it is also easy to 
check directly).
\end{proof}

\bibliography{comnet}

\end{document}